# Temperature dependence of on-state inter-terminal capacitances ($C_{gd}$ and $C_{gs}$) of SiC MOSFETs and frequency limitations of their measurements

Alexander Tsibizov[a], Roger Stark[a] and Ulrike Grossner[a]

[a] Advanced Power Semiconductor Laboratory, ETH Zurich, Physikstrasse 3, 8092 Zurich, Switzerland

## ABSTRACT

Inter-terminal capacitances (ITCs) have major influence on the dynamic performance of power SiC MOSFETs. Knowledge of the exact values for the ITCs is required in order to perform accurate and predictive compact model simulations of their dynamic performance. Since commercial SiC MOSFETs are capable of operating in a wide range of temperatures, it is important to know the values of ITCs in the whole temperature range of operation. Direct measurements of the ITCs with standard equipment is possible only at low current levels (i.e. in the off-state ($V_{gs} < V_{th}$) for $V_{ds} > 0$ V), however their values in the on-state ($V_{gs} > V_{th}$) also influence the MOSFETs switching performance. In this work, ITCs of a planar SiC MOSFET in the on-state are studied by the means of a calibrated TCAD model, revealing substantial temperature dependence in the range of 300-450 K. In the first approximation, this temperature dependence of the ITCs can be explained by a weaker temperature dependence of the MOSFET channel resistance in comparison to its JFET and epitaxial layer resistances. In addition, it is shown that at high frequencies stray inductances of the TO-247-3 package result in a change of the extracted values of the on-state ITCs. This effect is already notable at 1 MHz.

## INTRODUCTION

Inter-terminal capacitances (ITCs) play important role in dynamic performance of the power SiC MOSFET [1,2]. Knowledge of values for the ITCs is required to perform accurate and predictive compact model simulations of their dynamic performance. An example of such a simple compact model (CM) is shown in Figure 1. Since commercial SiC MOSFETs are capable of operating in a wide range of temperatures, e.g. from -55 °C to 150 °C [3], it is important to know the parameters of the CM over the whole temperature range of device operation. However, at present often only the temperature dependence of the quasi-static MOSFET's current source and body diode are set in the CMs while ITCs do not depend on temperature, see e.g. [4]. Moreover, supported by the off-state measurements it has been suggested that the temperature dependence of power MOSFET's ITCs is weak [5]. As shown in [6], direct measurements of the ITCs with standard equipment is possible only at low current levels (i.e. in the off-state ($V_{gs} < V_{th}$) for $V_{ds} > 0$ V), since the power of commercial impedance measurement systems are limited much below the full dynamic range of power MOSFETs. However, the values in the on-state ($V_{gs} > V_{th}$) also influence the switching performance of the MOSFET [1]. Although a number of methods has been suggested recently for the on-state measurements of ITCs of SiC MOSFET [7,8,9], the on-state ITCs and in particular their temperature behavior remain poorly investigated. In this work, the on-state ITCs of a planar gate SiC MOSFET are studied by the means of a calibrated TCAD model [10], revealing a substantial temperature dependence in the range of 300-450 K.

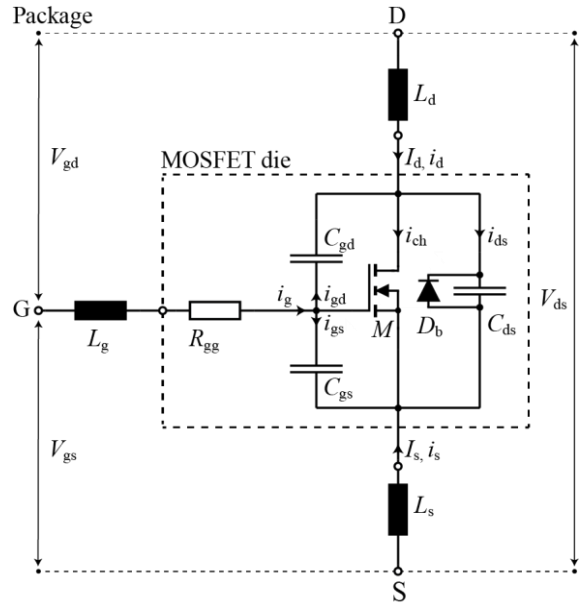

**Figure 1**. Electrical equivalent circuit of a power MOSFET including the inter-terminal capacitance $C_{gs}$, $C_{gd}$, $C_{ds}$, and internal gate resistance $R_{gg}$ and the parasitic inductances of the terminals $L_d$, $L_s$, $L_g$ of the package. The channel current source and the internal body diode of the MOSFET are represented by symbols M and $D_b$.

A frequency $f=1$ MHz is often used for the AC measurements of the ITCs in the off-state [3]. The same frequency $f=1$ MHz has been used in [7] for the on-state measurements of ITCs. Our TCAD simulations have shown that at high frequencies stray inductances of TO-247-3 package result in a change of extracted values of the on-state ITCs and this effect is quite pronounced already at 1 MHz.

## METHOD

$C_{gd}$ and $C_{gs}$ of a planar SiC MOSFET were extracted from small-signal AC mixed-mode simulations of a calibrated 2D TCAD model [10], aiming to reproduce electrical characteristics of a commercial device [3]. The stray inductances of the TO-247-3 package were included in a simplified manner following [11], see Figure 1, with the values $L_g$=8.2 nH, $L_d$=3.57 nH, $L_s$=5.89 nH, and the mutual coupling coefficients $K_{dg}$=0.294, $K_{ds}$=0.342, $K_{gs}$=0.121 extracted in ANSYS Q3D Extractor at 1 MHz. A shunt resistor with the same value $R_a$=10 Ohm was added in parallel to each terminal inductance in order to improve convergence. An internal gate resistance $R_{gg}$=3.5 Ohm (obtained from our measurements of the C2M0080120D MOSFET at 1 MHz) has been used in the simulation. The AC signal was applied to the package electrodes and correspondent apparent ITCs were extracted from the AC susceptance $B=j\omega C_{apparent}$ at different frequencies. The capacitances are extracted during the quasi-stationary $V_{ds}$ ramp for different constant $V_{gs}$ at two constant device temperatures 300 K and 450 K, revealing their temperature and frequency dependence.

## RESULTS & DISCUSSION

The resulting TCAD simulation shows only a small temperature dependence of the ITCs in the off-state as can be expected from previous works [5], see Figure 2. However, this is not valid for the on-state ITCs, see Figure 3, where the temperature dependence is much more pronounced. Such a difference between the temperature dependence of ITCs in the off-state and on-state can be explained with the help of simplified small signal equivalent circuits as presented in Figure 4. In the off-state, the gate oxide capacitances $C_{g1}$ and $C_{g2}$ are separated by a depleted region and charged correspondingly only from the source or the drain, thus correspondingly contributing only to $C_{gs}$ or $C_{gd}$. In the on-state, they are connected by the conducting channel, thus making the situation similar to the one considered in [12]. $C_{g1}$ and $C_{g2}$ can be charged through both source and drain contacts, therefore contributing to both $C_{gs}$ and $C_{gd}$. The share of this capacitance ($C_{g1}+C_{g2}$) attributed to $C_{gs}$ or $C_{gd}$ depends on the ratio $R_{ch}/(R_{jfet}+R_{epi})$. Hence, the temperature dependence of $C_{gs}$ and $C_{gd}$ is caused by the large differences between temperature dependence of the channel and drain resistances [13], whose ratio determines the attribution of charge in the channel and JFET region to the source or to the drain. Thereby, it determines the decomposition of $C_{gg}= C_{gs} + C_{gd}$ into $C_{gs}$ and $C_{gd}$. The explanation above is quite obvious for the on-state case of $V_{ds} \approx 0$, when $C_{gg}$ is approximately equal to the gate oxide capacitance $C_{gg} \approx C_{ov}+C_{g1}+C_{g2}$, and $C_{gg}(T=300 K) \approx C_{gg}(T=450 K)$. However, for larger $V_{ds}$ the values of the $C_{gg}$ capacitance of the double-diffused metal-oxide semiconductor (DMOS) device may exceed the gate-oxide capacitance [14]. This results from the variation on the voltage drop over the gate oxide which is larger than the change in the imposed gate bias when reaching the quasi-saturation region [14].

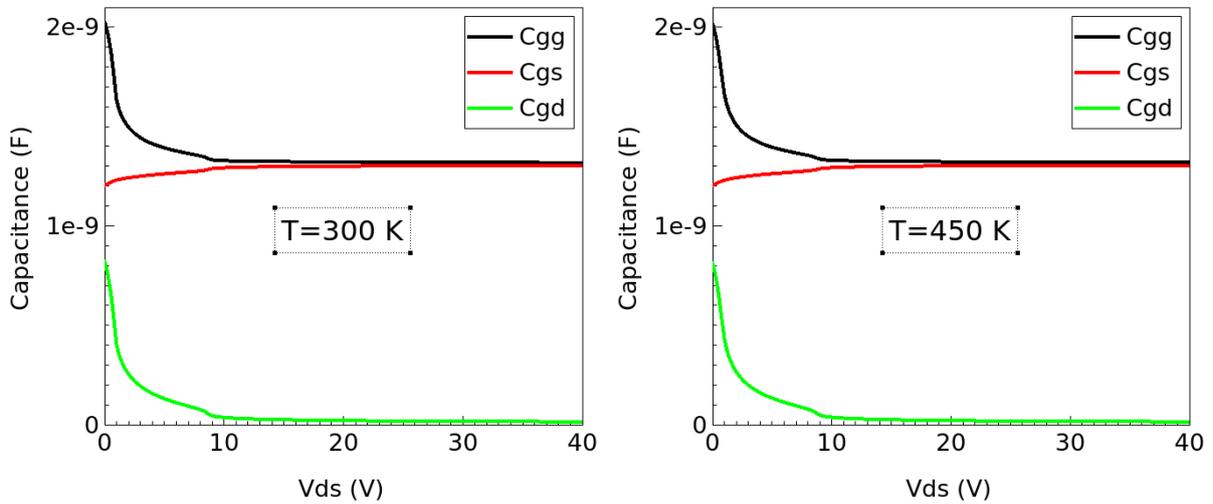

**Figure 2.** Simulated $C_{gg}$, $C_{gs}$ and $C_{gd}$ in the off-state ($V_{gs}=0$ V), $f=100$ kHz. Left plot T=300 K, right one T=450 K.

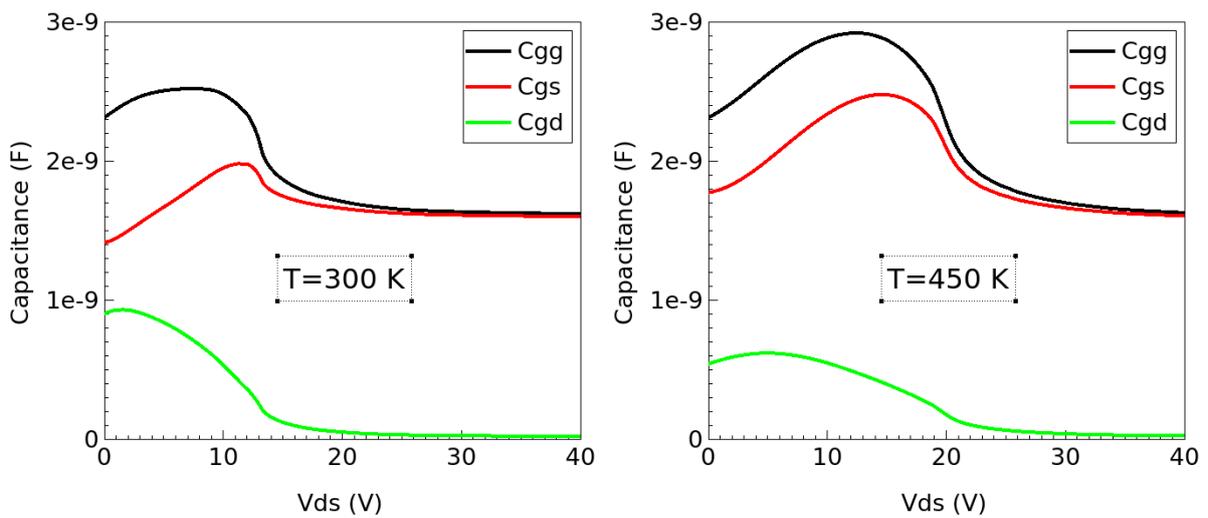

**Figure 3.** Simulated $C_{gg}$, $C_{gs}$ and $C_{gd}$ in the on-state ($V_{gs}=10$ V), $f=100$ kHz. Left plot T=300 K, right one T=450 K.

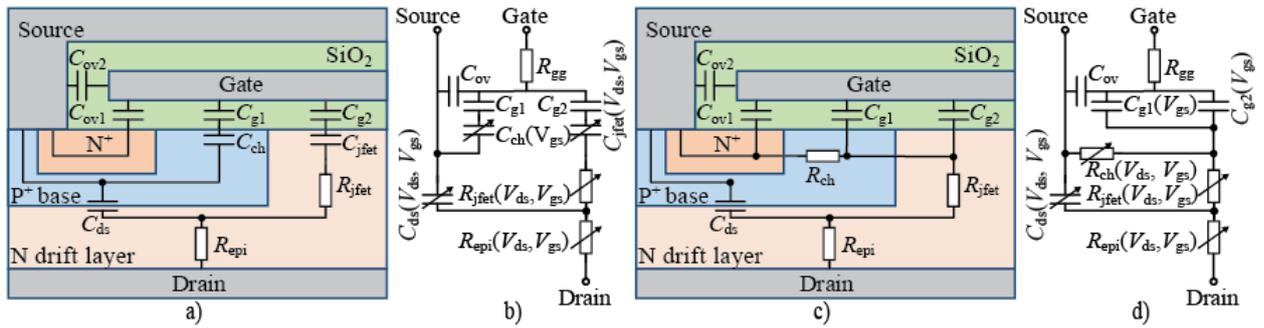

**Figure 4.** Simplified electrical equivalent circuit of a vertical double-diffused power MOSFET half-cell structure for a) a closed channel ($V_{gs} < V_{th}$), and c) open channel ($V_{gs} \gg V_{th}$). The corresponding small-signal equivalent circuits b) and d) indicate the voltage dependencies of the capacitances and resistances.

$V_{gs}$=10 V is used for the demonstration of the on-state ITCs because it is in the region of the Miller plateau voltage. However, it represents the overall behavior of the on-state ITCs well.

The frequency $f$=1 MHz is often used for the measurements of the ITCs in the off-state [3]. TCAD simulations with the stray inductance of the TO-247-3 package show little frequency dependence of the off-state ITCs up to 1 MHz, see Figure 5. A frequency $f$=1 MHz was used in [7] for the extraction of the apparent on-state ITCs of packaged devices. Figure 5 shows that in the on-state $f$=1 MHz is large enough to cause a noticeable difference of extracted apparent ITCs in comparison to the internal (related to the die design) ITCs due to the stray inductances of the TO-247-3 package. The internal ITCs correspond to the ones extracted at low frequency (f ≤ 100 kHz). Hence, for the real on-state measurements there is a contradiction in requirements: on the one hand, self-heating affects the extracted ITCs due to a change of device temperature, requiring smaller measurement duration to avoid it. On the other hand, if the measurement duration is too small, the influence of the package stray inductance will distort the results of the measurements, because smaller measurement durations require larger frequencies for AC measurements [7]. A similar problem exists also for the real on-state ITCs measurements based on the voltage sweep method [9]. Usage of a low stray inductance package could mitigate the problem.

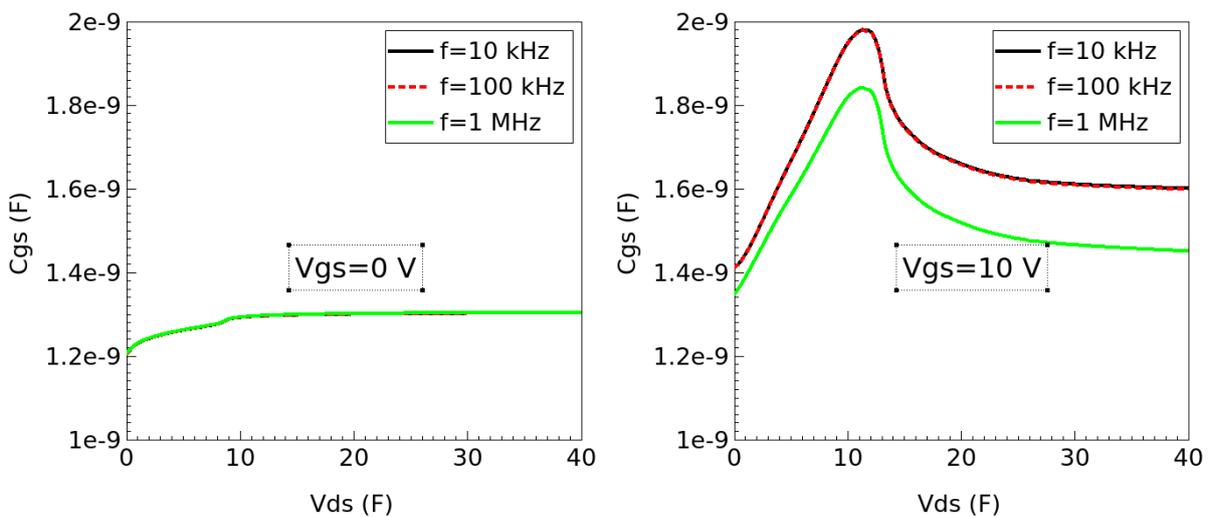

**Figure 5.** Simulated frequency dependence of apparent $C_{gs}(V_{ds})$ at T=300 K. Left plot: in the off-state ($V_{gs}$=0 V). Right plot: in the on-state ($V_{gs}$=10 V). Curves for $f$=10 kHz and $f$=100 kHz coincide.

# CONCLUSION

Unlike in the off-state, TCAD simulations of SiC MOSFETs show a noticeable temperature dependence of the on-state ITCs ($C_{gd}$ and $C_{gs}$) in the temperature range 300 K – 450 K. Therefore, such temperature dependence should be considered for accurate CM simulations of such devices.

In addition, the values of the ITCs' extracted from the on-state measurements of TO-247-3 packaged devices are strongly affected by the package stray inductances already at $f$=1 MHz. Using lower frequencies for the AC measurements of the on-state ITCs may lead to a strong self-heating (due to a longer measurement time). Thus, the usage of another package with smaller stray inductances for direct measurements of the internal ITCs related to the die design is suggested.

## Acknowledgments

Dr. Ivana Kovačević-Badstübner is gratefully acknowledged for providing the stray inductance parameters of TO-247-3 package.